%% file: Nb3Al_main.tex
\documentclass[%
 aip,
 amsmath,amssymb,
 reprint,%
]{revtex4-1}

\usepackage{graphicx}
\usepackage{dcolumn}
\usepackage{bm}
\usepackage[utf8]{inputenc}
\usepackage[T1]{fontenc}
\usepackage{mathptmx}
\usepackage{etoolbox}
\usepackage[english]{babel}
\usepackage[explicit]{titlesec}
\usepackage{soul}
\usepackage{xcolor}
\sethlcolor{yellow} 
\usepackage{tabularx, multirow, makecell}
\usepackage{float}

\titleformat{\section}  
  {\large\bfseries\sffamily}     
  {\thesection}         
  {1em}                
  {#1}      

\renewcommand{\thesection}{\Roman{section}.}

\titleformat{\subsection}
  {\normalsize\bfseries\sffamily}  
  {\thesubsection} 
  {1em}  
  {#1}  
  
\renewcommand{\thesubsection}{\Alph{subsection}.}

\makeatletter
\def\bbl@set@language#1{%
  \edef\languagename{%
    \ifnum\escapechar=\expandafter`\string#1\@empty
    \else\string#1\@empty\fi}%
  \@ifundefined{babel@language@alias@\languagename}{}{%
    \edef\languagename{\@nameuse{babel@language@alias@\languagename}}%
  }%
  \select@language{\languagename}%
  \expandafter\ifx\csname date\languagename\endcsname\relax\else
    \if@filesw
      \protected@write\@auxout{}{\string\select@language{\languagename}}%
      \bbl@for\bbl@tempa\BabelContentsFiles{%
        \addtocontents{\bbl@tempa}{\xstring\select@language{\languagename}}}%
      \bbl@usehooks{write}{}%
    \fi
  \fi}
\newcommand{\DeclareLanguageAlias}[2]{%
  \global\@namedef{babel@language@alias@#1}{#2}%
}
\makeatother

\DeclareLanguageAlias{en}{english}

\makeatletter
\def\@email#1#2{%
 \endgroup
 \patchcmd{\titleblock@produce}
  {\frontmatter@RRAPformat}
  {\frontmatter@RRAPformat{\produce@RRAP{*#1\href{mailto:#2}{#2}}}\frontmatter@RRAPformat}
  {}{}
}%
\makeatother

\addto\captionsenglish{\renewcommand{\figurename}{\textbf{Figure}}}
\renewcommand{\thefigure}{\textbf{\arabic{figure}}}

\addto\captionsenglish{}

\begin{document}

\preprint{AIP/123-QED}

\title{An evaluation of A15 Nb\textsubscript{3}Al superconducting thin films for application in quantum circuits}

\author{Joseph Falvo}
\affiliation{Department of Materials Science and Engineering, University of Maryland, College Park, Maryland 20740, USA}
\affiliation{Laboratory for Physical Sciences, University of Maryland, College Park, Maryland 20740 USA}

\author{Brooke Henry}
\affiliation{Department of Physics, University of Colorado, Boulder, Colorado 80309, USA}

\author{Bernardo Langa Jr.}
\affiliation{Laboratory for Physical Sciences, University of Maryland, College Park, Maryland 20740 USA}
\affiliation{Department of Physics, University of Maryland, College Park, Maryland 20740, USA}

\author{Rohit Pant}
\affiliation{Laboratory for Physical Sciences, University of Maryland, College Park, Maryland 20740 USA}
\affiliation{Department of Physics, University of Maryland, College Park, Maryland 20740, USA}

\author{Ashish Alexander}
\affiliation{Laboratory for Physical Sciences, University of Maryland, College Park, Maryland 20740 USA}
\affiliation{Department of Physics, University of Maryland, College Park, Maryland 20740, USA}

\author{Jason Dong}
\affiliation{Laboratory for Physical Sciences, University of Maryland, College Park, Maryland 20740 USA}
\affiliation{Department of Physics, University of Maryland, College Park, Maryland 20740, USA}

\author{Kasra Sardashti}
\altaffiliation{Corresponding author: ksardash@umd.edu}
\affiliation{Laboratory for Physical Sciences, University of Maryland, College Park, Maryland 20740 USA}
\affiliation{Department of Physics, University of Maryland, College Park, Maryland 20740, USA}

\date{\today}

\begin{abstract}
A15 superconductors are distinguished by their high critical temperatures, magnetic fields, and current-carrying capabilities. Among them, Nb\textsubscript{3}Al is of particular interest for superconducting quantum circuits as a means to extend device operating temperatures, provided that its electrodynamic properties are well understood. Here, we report on the synthesis of Nb\textsubscript{3}Al thin films by magnetron co-sputtering followed by rapid thermal processing, yielding superconducting transition temperatures above 16 K. Microwire devices patterned from these films exhibit a coherence length of 3.2 nm and superfluid densities as low as $1.1 \times 10^{26}$ m\textsuperscript{-3}, suggesting that Nb\textsubscript{3}Al may enable high kinetic inductance in thinner films. Coplanar waveguide resonators fabricated on Nb\textsubscript{3}Al demonstrate single-photon internal quality factors up to $2.26 \times 10^5$. These results establish Nb\textsubscript{3}Al as a promising material platform for the development of superconducting quantum circuits operating at elevated temperatures, contingent on appropriate control of interfacial chemistry and surface morphology.
\end{abstract}

\maketitle

\section{\label{sec:level1}Introduction}
Engineering materials to enhance fault tolerance is key to realizing scalable superconducting (SC) qubit architectures for quantum computing.\cite{richardson_materials_2020} Recent developments in qubit technology have highlighted the need for superconducting materials with higher critical temperatures (\textit{T\textsubscript{c}}) and critical magnetic fields (\textit{B\textsubscript{c}}) than those offered by conventional choices such as niobium and aluminum. This need is driven by the growing interest in hybrid superconductor-semiconductor, fluxonium, and high-temperature qubit platforms operating above 300 mK, all of which require superconductors that are resilient to external perturbations.\cite{langa_solid-state_2024,nguyen_high-coherence_2019,siddiqi_engineering_2021,anferov_superconducting_2024} Additionally, using materials with higher critical temperatures could reduce quasiparticle-induced loss, a major source of decoherence in SC qubits.\cite{serniak_hot_2018} Compound superconductors have been explored as promising candidate materials for such applications. Of these, NbTiN and NbN have been of particular interest due to their high \textit{T\textsubscript{c}} ($\leq 17$K), \textit{B\textsubscript{c}} ($\leq 22$ T), and kinetic inductance ($ L_k\leq 200$ pH/$\square$ for films with thicknesses  $\leq13$ nm). These properties have enabled their successful integration into a range of quantum devices, from traveling wave parametric amplifiers to tunable filters/resonators.\cite{muller_magnetic_2022,kim_field-resilient_2025,yang_kinetic_2024,giachero_characterization_2023,shiino_improvement_2010,sidorova_magnetoconductance_2021,frasca_nbn_2023} However, these materials have been shown to suffer from issues such as phase impurities and nitrogen non-uniformity that can impair these favorable superconducting properties, and thus compromise device performance.\cite{nulens_catastrophic_2023,supple_atomic_2024} Intermetallic compounds such as A15 superconductors can present a promising alternative for quantum device applications where high \textit{T\textsubscript{c}}, \textit{B\textsubscript{c}}, and \textit{L\textsubscript{k}} are favorable. Unlike disordered nitrides, their enhanced kinetic inductance arises from strong electron-phonon coupling and effective mass enhancement, thereby mitigating dissipation linked to disorder and inhomogeneity.\cite{arko_haas-van_1978,terashima_haasvan_2001,dew-hughes_superconducting_1975} 

Intermetallic compounds with A15 crystal structure (representative formula A\textsubscript{3}B, space group: Pm$\bar{3}$n) were extensively studied between the 1950s and 1970s as high-temperature superconductors. Research interest declined following the discovery of cuprate superconductors, whose \textit{T\textsubscript{c}}’s far exceed those of the A15 family.\cite{dew-hughes_superconducting_1975} A wide range of binary A15 compounds were synthesized in bulk crystal, wire, and thin film form factors with critical temperatures reaching up to 23.2 K.\cite{testardi_superconductivity_1974} Given that more than ten A15 compounds exhibit higher \textit{T\textsubscript{c}} and \textit{B\textsubscript{c}} than conventional superconductors such as aluminum and niobium (e.g., Nb\textsubscript{3}Al, Nb\textsubscript{3}Ge, Nb\textsubscript{3}Sn, V\textsubscript{3}Si), these materials merit renewed exploration as candidates for superconducting quantum electronics. However, significant gaps remain in the cryogenic electrical characterization of A15 thin films, with key parameters such as coherence length, mean free path, London penetration depth, and Cooper pair density yet to be determined for many compounds. More importantly, even in cases where direct current (DC) properties have been studied, data on microwave loss of A15 thin films at milikelvin (mK) temperatures are still lacking, limiting a comprehensive evaluation of their potential for their integration into SC qubits.\cite{moore_energy_1979,kwo_superconducting_1981,moore_superconductive_1976,rudman_microscopic_1984,tanabe_josephson_1984,tanabe_nbbased_1987,tanabe_nb3aloxidepb_1983,tanabe_nbbased_1985}

A15 Nb\textsubscript{3}Al is of particular interest, exhibiting thin film critical temperatures above 16 K and  bulk critical magnetic field exceeding 30 T.\cite{tanabe_selfepitaxial_1984,foner_effects_1981} In the past, Nb\textsubscript{3}Al received attention as a promising material for cryoelectronics due to its potential to maintain large current densities and high SC gap ($>$2.15 meV).\cite{kwo_superconducting_1981,tanabe_nbbased_1987,tanabe_nbbased_1985} The first thin films were realized through electron-beam coevaporation, with \textit{T\textsubscript{c}}’s reaching up to 16.5 K.\cite{kwo_superconducting_1981,kwo_nb3al_1980} However, the most common growth method to date has been DC magnetron sputtering from a fixed near-stoichiometric Nb\textsubscript{3}Al target onto a heated substrate held between 600-800  $^\circ$C, with the best films achieving \textit{T\textsubscript{c}}’s up to 17.7 K.\cite{tanabe_nbbased_1987,tanabe_selfepitaxial_1984,dahlgren_high-rate_1975,tanabe_properties_1982,tanabe_composition_1982} More recently, films with \textit{T\textsubscript{c}}’s of 15.7 K have also been achieved using cosputtering of Nb and Al targets onto a heated substrate, followed by rapid thermal processing (RTP) at 1000 $^\circ$C.\cite{dochev_nb3al_2008} The A15 phase is metastable, making the structural and superconducting properties of films very sensitive to growth conditions, particularly deposition temperature and RTP conditions. An advantage of Nb\textsubscript{3}Al is its relative stability compared to the other A15s. Single-phase A15 exists within a composition range of 17.5-21.5 at. \% Al at 1000 $^\circ$C in bulk samples, but can persist in the range of 9-30 at. \% Al.\cite{jorda_new_1980} Compositional studies in thin films and Josephson junctions (JJs) support this, with maximal \textit{T\textsubscript{c}}’s and the best measured junction characteristics occurring within the single-phase range.\cite{kwo_superconducting_1981,dahlgren_high-rate_1975} 

While Nb\textsubscript{3}Al films have previously been integrated into JJs, they were used primarily as a platform to probe superconducting properties rather than for superconducting circuit applications. Those early devices employed a single A15 Nb\textsubscript{3}Al electrode, a native oxide or amorphous silicon barrier, and a Pb counterelectrode, and no microwave-frequency characterization was performed, leaving their relevance to modern SC circuits unclear.\cite{kwo_superconducting_1981,tanabe_nbbased_1987,tanabe_nb3aloxidepb_1983,tanabe_nbbased_1985} Here, we explore the A15 Nb\textsubscript{3}Al system through the lens of SC microwave circuit applications. We characterize the superconducting and cryogenic microwave properties of Nb–Al thin films, motivated by the material’s high \textit{T\textsubscript{c}} and \textit{B\textsubscript{c}}, as well as the ability to engineer alloy disorder to tune Cooper-pair density and therefore kinetic inductance. In this work, we grow thin films of superconducting Nb-Al alloys via direct current (DC) and radio-frequency (RF) co-sputtering under high vacuum (HV), achieving critical temperatures up to 16.64 K. Microwave measurements reveal a critical current density (\textit{J\textsubscript{c}}) of $3.2 \times 10^6$ A cm\textsuperscript{-2} and superfluid density on the order of $1.1 \times 10^{26}$ m\textsuperscript{-3}, comparable to NbN, indicating similar \textit{L\textsubscript{k}}. Resonator devices exhibit a median single-photon internal quality factor (\textit{Q\textsubscript{i}}) of $2.26 \times 10^5$ between 4.5-5.25 GHz, demonstrating the material’s compatibility with SC microwave circuitry. Our combined materials and device characterization  studies highlight the opportunities and challenges in integrating A15 thin films into superconducting qubit devices.

\begin{table*}
    \centering
    \caption{\label{tab:table1} Sample labels and corresponding processing parameters including substrate angle, deposition temperature, and RTP conditions. Aluminum content and estimated thicknesses are also listed for each film.}
    \begin{tabular*}{\textwidth}{@{\extracolsep{\fill}} c c c c c c c c}
    \hline
         
    Sample Group &
    \makecell{Substrate\\Angle ($^\circ$)} &
    \makecell{Deposition\\Temperature ($^\circ$C)} &
    \makecell{Al Content\\(at. \%)} &
    \makecell{Estimated\\Thickness (nm)} &
    \makecell{RTP\\Condition} &
    Sample Code &
    \makecell{RMS\\Roughness (nm)}\\
    \hline

    L1 & 10 & RT & 25 & 225 &
    \makecell[c]{None \\ Ar flow \\ Vacuum} &
    \makecell[c]{L1 \\ L1-AR \\ L1-VAC} &
    \makecell[c]{2.05 \\ 7.43 \\ 4.72} \\
    \hline
  
    H1 & 10 & 650 & 24 & 215 &
    \makecell[c]{None \\ Ar flow \\ Vacuum} &
    \makecell[c]{H1 \\ H1-AR \\ H1-VAC} &
    \makecell[c]{1.27 \\ 8.04 \\ 4.06} \\
    \hline

    H2 & 5 & 700 & 25 & 210 &
    \makecell[c]{None \\ Ar flow \\ Vacuum} &
    \makecell[c]{H2 \\ H2-AR \\ H2-VAC} &
    \makecell[c]{0.66 \\ 14.06 \\3.58} \\
    \hline
    \end{tabular*}
\end{table*}

\section{Experimental}
\subsection{Thin film growth}
Films investigated in this work were deposited using DC-RF magnetron co-sputtering, utilizing pure Nb and Al targets, driven by DC and RF power supplies, respectively, and at a fixed angle relative to each other. The angle between the sources and the substrate was changed via substrate rotation about the epicenter of the chamber, enabling compositional variation without changing the sputter conditions at each source. In addition to substrate angle variations, substrate temperature was varied during the deposition process. Chamber pressure was maintained at 3 mTorr by actively controlled Ar flow. The target thickness of all deposited films was 300 nm. Prior to deposition, all substrates underwent cleaning by sonication in sequential baths of acetone, methanol, and isopropanol for 2 minutes, followed by dry nitrogen gas to remove solvents before the formation of evaporative residues could occur. This cleaning process is referred to hereafter as AMI cleaning. The substrates used were Si(001), which enabled elemental analysis of the films while excluding substrate signal, and Al\textsubscript{2}O\textsubscript{3}(0001), which allowed for high-temperature processing without formation of silicides.

After deposition, samples underwent rapid thermal processing (RTP) where they were placed in a SiC-coated graphite susceptor and heated to 1000 $^\circ$C via IR radiation. The variables explored were hold time at the maximum temperature and the flow rate of Ar gas through the system. Hold time was varied between 30 and 120 seconds, and Ar flow rate set to 200 sccm. For vacuum processing, the RTP chamber was pumped continuously by a turbomolecular pump with typical base pressures of $5 \times 10^{-5}$ mbar. All RTP runs had average heating rates of 6 $^\circ$C/s and average cooling rates of 2 $^\circ$C/s down to 400 $^\circ$C.  Several RTP conditions were tested for sample L1, and the two which gave the highest $T_c$ were chosen as standard recipes for evaluating other sample groups.

\subsection{Microfabrication}
After RTP, select wafers were selected based on \textit{T\textsubscript{c}} for fabrication of devices to enable additional DC and microwave property characterization. These samples were patterned into micron-scale wires (microwires) and coplanar waveguide (CPW) resonators with frequencies between 4.5 and 5.5 GHz. The nanofabrication process utilized photolithography followed by an inductively-coupled plasma reactive ion etch (ICP-RIE) process using BCl\textsubscript{3} and Cl\textsubscript{2} gases. Etch time was chosen to fully etch through the deposited film, isolating the Nb\textsubscript{3}Al devices from one another. Resonators and microwires were simultaneously fabricated from the same chip so that properties could be compared directly.

\subsection{Characterization}
Chemical composition was first measured  by energy dispersive x-ray spectroscopy (EDS) at 5 kV to minimize the interaction volume within the substrate. EDS on films deposited on Si substrates was used to determine the Nb:Al ratio, avoiding the substrate-related artifacts observed with sapphire. Composition was later assessed independently by Rutherford backscatter (RBS) experiments on sapphire substrates and was found to be consistent with the EDS results. X-ray photoelectron spectroscopy (XPS) was also performed to evaluate surface composition of the films before and after RTP. The XPS uses a monochromatic Al X-ray source and a 100 \textmu m beam spot size. In-situ Ar ion milling within the XPS system was utilized for depth profiling measurements on the samples.

Structural characterization was performed using thin-film X-ray diffraction (XRD) with a two-theta range of 20-100 degrees. XRD spectra were compared to existing libraries to determine the phases present within the film. The surface morphology of each film was also studied using atomic force microscopy (AFM). The AFM uses a silicon tip in tapping mode to scan the sample surface over 5 \textmu m $\times$ 5 \textmu m regions.

Magnetotransport measurements were performed using low-frequency AC excitation from a lock-in amplifier in a van der Pauw (VdP) four-wire configuration within a PPMS, enabling temperature control from 1.8–300 K and magnetic fields up to 14 T. Critical current density in microwires was determined using a DC current source to supply the bias current while a lock-in amplifier measured the differential resistance. The superconducting coherence length, $\xi_0$, was extracted from systematic magnetotransport measurements by sweeping the magnetic field from 0 to 14 T at multiple temperatures near \textit{T\textsubscript{c}}.

Microwave loss measurements under typical SC qubit operational conditions (i.e., mK temperatures and 4-8 GHz frequencies) were conducted  on devices with  CPW resonators fabricated from Nb\textsubscript{3}Al on sapphire. A chip containing five resonators was wedge bonded to a copper printed circuit board (PCB). The PCB and chip were then mounted in an Al package with SMA connectors soldered to the feedline of the PCB. The package was then cooled down in a dilution refrigerator with a base temperature of 10 mK. A vector network analyzer (VNA) was used in series with a variable attenuator to pass microwave frequency signals through the wiring of the refrigerator and measure the real and imaginary components of the output signal from the device. See Supporting Information, \textbf{Figure S1} for a complete diagram of the RF circuit and optical images of the resonator chip. The input power and operating temperature were systematically varied to reveal the dominant loss mechanisms.\cite{alexander_power_2025,sage_study_2011}

\section{Results and Discussion}
Samples highlighted in the following experiments and discussion were deposited under a variety of substrate angles and temperatures and underwent RTP under multiple conditions. For ease of reference, the deposition and RTP conditions for each sample are detailed in \textbf{Table} \ref{tab:table1}. The substrate angle refers to the angle between the substrate surface normal and the vertical axis of the deposition chamber, with $0^\circ$ corresponding to the surface normal lying equidistant between the two targets. Angles greater than $0^\circ$ indicate the substrate is tilted toward the Nb target during deposition. Varying deposition angle had a slight impact on Al content, but within a small range had minimal impact on deposition rate, with films ranging from 210 to 225 nm for equal deposition times. Deposition temperatures range from room temperature (RT) for group L1 (L denoting low-temperature growth) to 650 \textsuperscript{o}C and 700 \textsuperscript{o}C for groups H1 and H2, respectively (H denoting high-temperature growth).  Within each group, three sample variants are included: (i) \textbf{L1/H1/H2}, representing the as-deposited films; (ii) samples labeled \textbf{-AR} (e.g., L1-AR), denoting RTP in Ar at a constant flow of 200 sccm; and (iii) samples labeled \textbf{-VAC} (e.g., L1-VAC), denoting vacuum RTP carried out using a turbomolecular pump, with a base pressure of $1 \times 10^{-5}$ mbar prior to heating. The impact of RTP on surface morphology of the films is captured in the root-mean-square (RMS) roughness from AFM measurements, also listed in \textbf{Table} \ref{tab:table1}.

\begin{figure*}[t]
  \centering
  \includegraphics[scale=0.75]{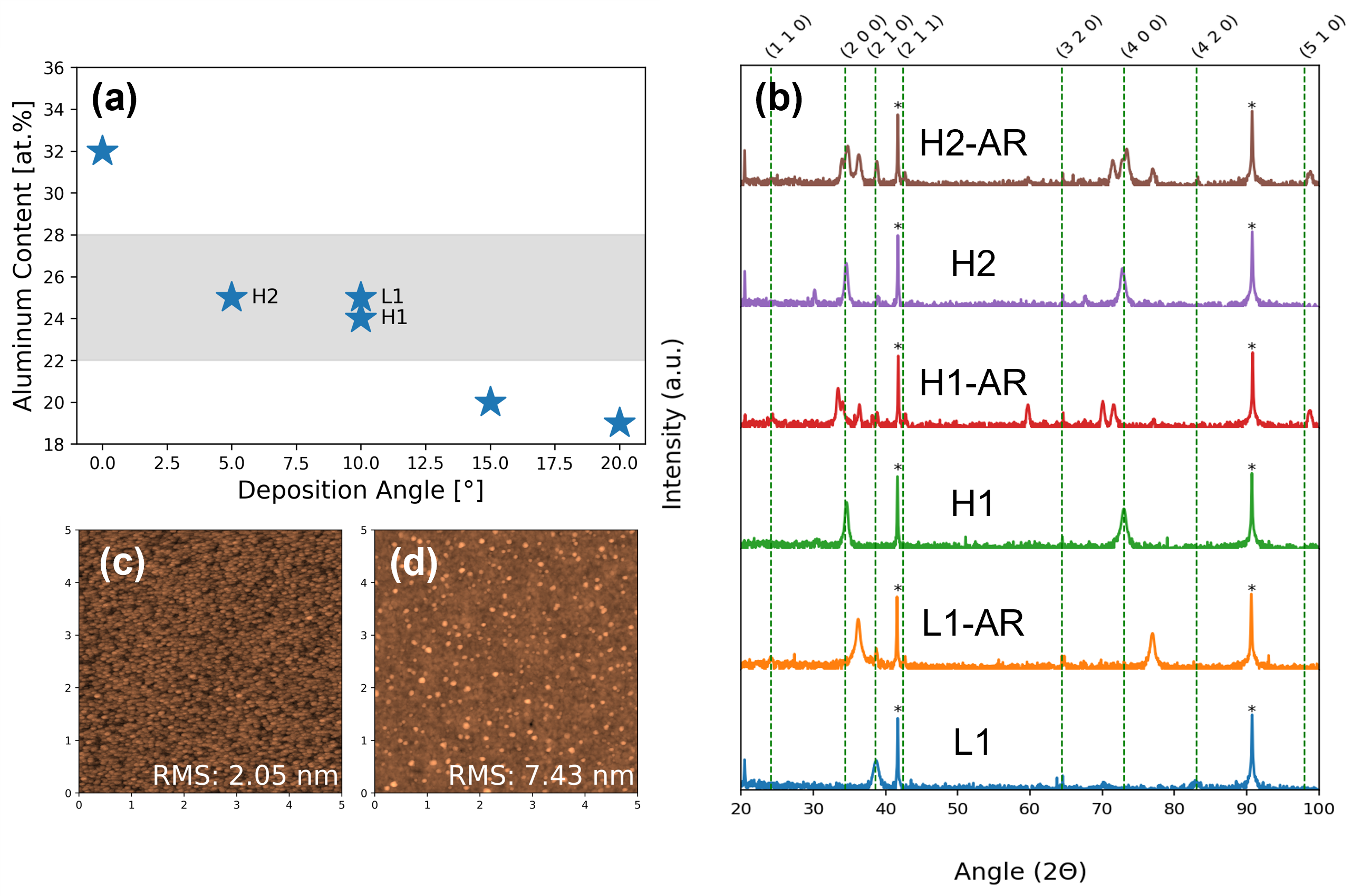}
  \caption{\textbf{Chemical and structural characteristics of co-sputtered Nb-Al thin films:} (a) Aluminum content in the Nb-Al thin films, expressed in atomic percent (at. \%), as a function of substrate angle during co-deposition. The gray shading highlights a region within 3 at. \% of stoichiometric 3:1 Nb:Al ratio. (b) XRD patterns for select samples where green dashed lines indicate expected peak positions from Nb\textsubscript{3}Al in the A15 structure, and * indicates peaks resulting from the sapphire (0001) substrate. (c) and (d) Surface topography maps for sample L1 before and after RTP under argon, respectively. RMS roughness shown in bottom right corner.}
  \label{fig:fig1}
\end{figure*}

\textbf{Figure \ref{fig:fig1}(a)} displays the composition – measured via RBS – of as-deposited Nb-Al thin films deposited at multiple substrate angles and temperatures. All films represented in \textbf{Figure \ref{fig:fig1}(a)} were between 210 and 225 nm in thickness. The gray region in the plot highlights conditions within 3 at. \% of the stoichiometric 3:1 Nb:Al ratio. Though A15 Nb\textsubscript{3}Al is stable at 3:1 Nb:Al at temperatures well above 1000 $^\circ$C, below that temperature the stability range narrows and shifts to Al-poor compositions. At room temperature, the A15 structure is stable from 17.5 to 24.5 at. \% Al, which overlaps the highlighted region. This figure highlights that we are able to control the film composition from Al-rich to Al-poor by simply varying the substrate angle while maintaining the overall sputtering dynamics via pressure or power density. The tunability provided by this approach allowed for verification that the maximum $T_c$ for Nb-Al films occurred in the grey zone ($22-28$ at. \% of Al) corresponding to deposition angles of $5^\circ - 10^\circ$.  Thereby, the remainder of this paper focuses on  the characterization of sample groups L1, H1, and H2 with compositions that are in the optimal A15 window (see \textbf{Table \ref{tab:table1}}).

The XRD patterns in \textbf{Figure} \ref{fig:fig1}\textbf{(b)} indicate that the co-sputtering process can produce films with the A15 structure even at low substrate temperatures. Sample L1 shows a dominant (210) reflection, accompanied by a weaker (420) peak, suggesting a preferred (210) orientation for films grown at room temperature. For samples H1 and H2, the diffraction is dominated by the (200) and (400) reflections, indicating that increasing substrate temperature promotes a reorientation of grains toward the (200)/(400) out-of-plane texture. Additional diffraction peaks appear after RTP under Ar (for samples L1-AR, H1-AR, and H2-AR), which are attributed to the introduction of Si and C into the films during the thermal processing step. Two of the secondary phases that form as a result of contamination during RTP are Nb\textsubscript{3}Si and Nb\textsubscript{6}AlSi, which share the A15 structure with a different lattice spacing compared to Nb\textsubscript{3}Al, resulting in the appearance of additional peaks. Shifts were also observed in both A15-related and substrate peaks. These shifts suggest changes in lattice spacing, which may arise from thermal effects, compositional variations, or stress relaxation during RTP.  A complete set of XRD spectra that includes data for L1-VAC, H1-VAC, and H2-VAC can be found in Supporting Information, \textbf{Figure S2}.

AFM surface topography maps for samples L1 And L1-AR  are shown in \textbf{Figure \ref{fig:fig1}(c)} and \textbf{(d)}, respectively. L1 had an RMS roughness of 2.05 nm for a film thickness of 225 nm, while L1-AR had RMS roughness of 7.43 nm. This increase in RMS roughness is due to the appearance of randomly distributed regions that protrude from the surface. This change suggests that RTP impacts near-surface composition and/or grain structure within the film. Surface topography maps for all samples can be found in Supporting Information, \textbf{Figure S3}. RMS roughnesses for all samples are listed in \textbf{Table \ref{tab:table1}}. For as-deposited samples, the RMS roughness was found to decrease with increased substrate temperature, consistent with increased adatom mobility during deposition. After RTP, the RMS roughness consistently increased. For RTP in vacuum, the resultant surface roughness was generally due to pitting of the surface, while the surface resulting from RTP under Ar flow showed surface protrusions. Based on the observed trends, it is likely that the surface morphology is changed due to the diffusion of Al towards the surface during RTP. The Al is removed from the surface via vacuum RTP, but collects and remains on the surface when RTP takes place in an Ar environment.

\begin{figure}[t]
  \centering
  \includegraphics[scale=0.52]{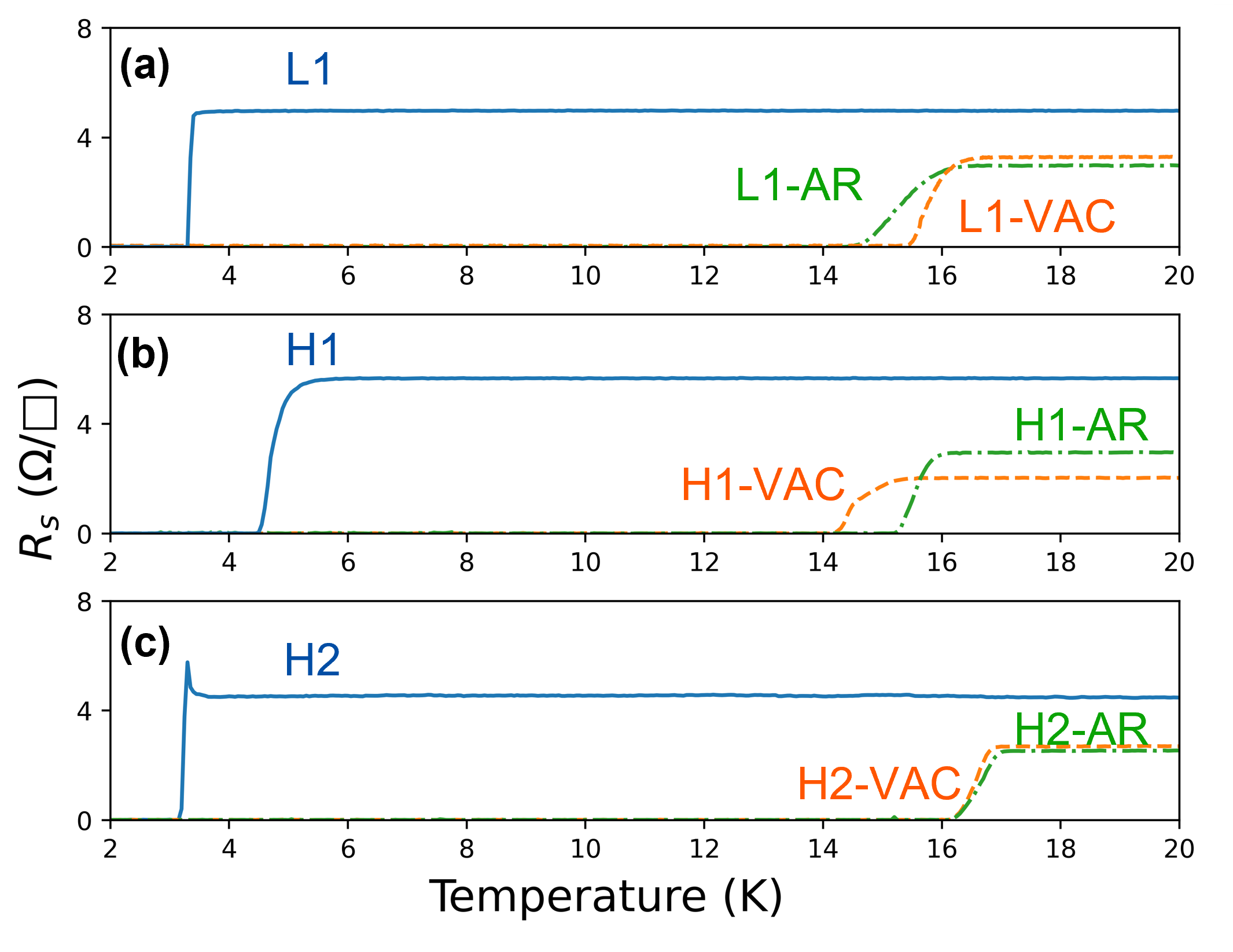}
  \caption{\textbf{Superconducting-normal transition in L1, H1, and H2 sample groups:} Sheet resistance vs temperature for samples derived from depositions (a) L1, (b) H1, and (c) H2 measured using a van der Pauw geometry. In each panel, the as-deposited sample (blue lines) is compared with samples treated by RTP under vacuum (-VAC, orange lines), and in Ar flow (-AR, green lines).}
  \label{fig:fig2}
\end{figure}

\textbf{Figure \ref{fig:fig2}} shows sheet resistance as a function of temperature for all samples, measured using a VdP configuration on 5 mm $\times$ 5 mm square pieces. \textbf{Figure \ref{fig:fig2}(a)} shows L1, L1-VAC, and L1-AR, and \textbf{Figure \ref{fig:fig2}(b)} and \textbf{(c)} show H1, H1-VAC, H1-AR and H2, H2-VAC, H2-AR, respectively. The highest \textit{T\textsubscript{c}} observed for the as-deposited sample (i.e., L1, H1, and H2) was 4.7 K despite the presence of A15 peaks in the XRD patterns shown in \textbf{Figure \ref{fig:fig1}(b)}. The suppression of \textit{T\textsubscript{c}} is likely due to A15 phases appearing as isolated regions in as-deposited films. After RTP, however, \textit{T\textsubscript{c}} was found to increase dramatically up to a maximum of 16.64 K measured at the midpoint. This is consistent with formation and expansion of A15 phases via a solid-state diffusion process across the films which remains metastable thanks to the RTP’s rapid cooling capabilities. The increase in \textit{T\textsubscript{c}} after RTP is accompanied by decrease in normal resistance of the films likely due to increase in grain size and the resulting decrease in grain boundary scattering.

RTP was also found to increase the width of the superconducting transition ($\Delta T_c = T_{R=0.9R_n} - T_{R=0.1R_n}$) and the residual resistance ratio (RRR = $R_{300 K}/R_{20 K}$). $T_c$, $\Delta T_c$, and RRR for all nine samples are listed in \textbf{Table \ref{tab:table2}}. The increase in $\Delta T_c$ indicates an overall increase in disorder of the films, due to a combination of compositional disorder from secondary phases (e.g., Nb\textsubscript{3}Si, Nb\textsubscript{2}C, Nb\textsubscript{6}AlSi) and structural disorder from Al segregation to the surface and grain boundaries. Across multiple high-temperature depositions, RTP in Ar was found to yield higher $T_c$ than RTP in vacuum, though RRR was slightly higher after vacuum RTP. The $T_c$ of samples from the H2 group, which contained 25 at. \% Al, did not vary as strongly between the RTP conditions. Moreover, sample H2-AR exhibited the highest $T_c$ of all samples, as well as reasonable $\Delta T_c$ and RRR. Because of these factors, H2-AR was chosen for further characterization in microfabricated devices.

\begin{table}[t]
    \centering
    \caption{\label{tab:table2} Electrical and superconducting  properties of sample groups L1, H1, and H2, including critical temperature ($T_c$), width of S-N transition ($\Delta T_c$), and residual resistance ratio (RRR).}
    \begin{tabular*}{\columnwidth}{@{\extracolsep{\fill}} l c c c}
        \hline
        Sample Code & $T_c$ (K) & $\Delta T_c$ (K) & RRR \\
        \hline
            L1 & 3.31 & 0.08 & 1.02 \\
            L1-VAC & 15.81 & 0.67 & 1.42 \\
            L1-AR & 15.20 & 1.17 & 1.37 \\
            H1 & 4.70 & 0.46 & 1.02 \\
            H1-VAC & 14.49 & 0.75 & 1.74 \\
            H1-AR & 15.55 & 0.52 & 1.39 \\
            H2 & 3.23 & 0.06 & 1.21 \\
            H2-VAC & 16.55 & 0.46 & 1.52 \\
            H2-AR & 16.64 & 0.57 & 1.47 \\
        \hline
    \end{tabular*}
\end{table}

\begin{figure}[b]
  \centering
  \includegraphics[scale=0.75]{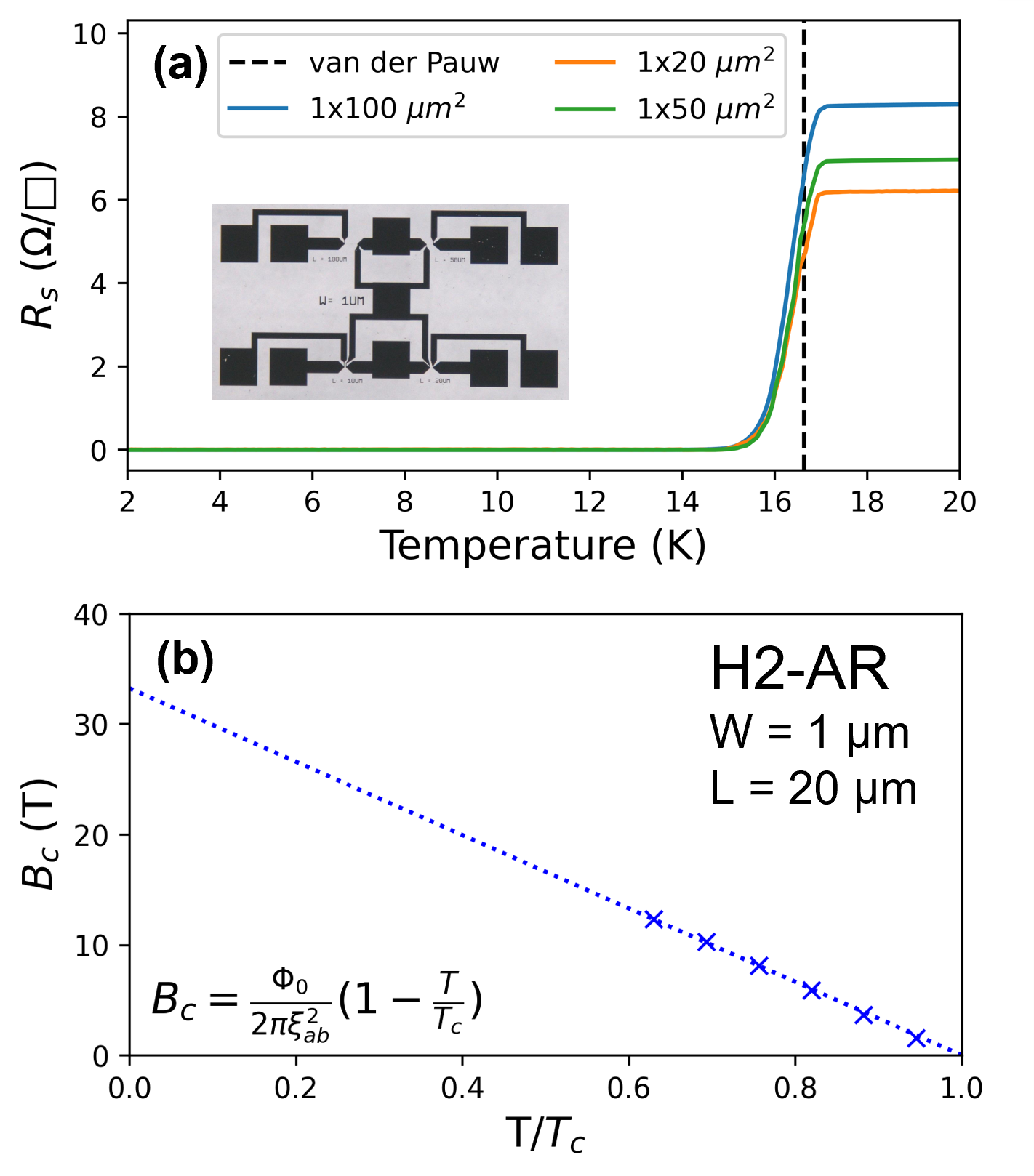}
  \caption{\textbf{Superconducting characteristics of H2-AR microwires:} (a) Sheet resistance vs temperature curves for a set of microwires fabricated from H2-AR (inset: optical micrograph of the microwire test device). Vertical dashed line indicates $T_c$ of H2-AR measured in VdP configuration. (b) Critical magnetic field ($B_c$) vs temperature normalized to $T_c$ for a typical microwire, which is fit to the GL relation for 2D superconductors.}
  \label{fig:fig3}
\end{figure}

\textbf{Figure \ref{fig:fig3}(a)} shows the sheet resistance vs temperature for three representative microwires fabricated on an H2-AR thin film with widths of 1 \textmu m and lengths of 20, 50, and 100 \textmu m. The vertical dashed line shows the \textit{T\textsubscript{c}} of the H2-AR film measured in the VdP geometry. An optical image of the microwires measured in this study are shown in the inset of \textbf{Figure \ref{fig:fig3}(a)}. This geometry was designed to enable simultaneous 4-point measurement (unique I+, I-, V+, V-)  of 4 distinct microwires. Upon narrowing to 1 \textmu m width, the \textit{T\textsubscript{c}} remained the same and showed virtually no dependence on the length of the microwire. \textit{T\textsubscript{c}} for all microwires are given in \textbf{Table \ref{tab:table3}}, along with RRR when measured. \textit{T\textsubscript{c}} and RRR were not found to be dependent on the microwire geometry, which confirms that at the micron scale, there are not significant compositional or structural variations in the films that lead to variations in the superconducting behavior. Therefore, for the analysis that follows, it is assumed that the wires are representative of the base material. Differences in normal sheet resistance are due to variances in microwire geometry away from nominal dimensions caused by feature sizes close to the resolution limit of the microfabrication process.

\textbf{Figure \ref{fig:fig3}(b)} shows a typical magnetic field dependence for H2-AR microwire samples measured on a device with W $=1$ \textmu m and L $=20$ \textmu m. The behavior fits to the GL relation for a 2-D superconductor (dotted line), given by Equation \ref{eq1}, where \textit{B\textsubscript{c}} is the upper critical field, $\Phi _0$ is the superconducting flux quantum, and \textit{T} is the sample temperature.

\begin{equation}
    B_c(T)=\frac{\Phi_0}{2\pi \xi_0^2}(1-\frac{T}{T_c})
    \label{eq1}
\end{equation}

As can be seen from Equation \ref{eq1}, the upper critical field $B_c$ varies linearly with temperature in 2D superconductors, in contrast to three-dimensional systems where $B_c \propto T^2$. Although the films in this study exceed 200 nm in thickness, well above both the coherence length ($\xi_0 \sim3$ nm) and the penetration depth, the 2D model provided a significantly better fit than a 3D description. The success of the 2D framework is unexpected for such thick films; however, the linear dependence may reflect anisotropy arising from microstructural or electronic inhomogeneity, effectively restricting superconducting transport to quasi-2D domains. From the fit, we obtained \textit{B\textsubscript{c}(0 K)} = 33.2 T and $\xi_0$ = 3.1 nm, values that closely match those from VdP samples grown under identical conditions, \textit{B\textsubscript{c}(0 K)} of 32.6 T and $\xi_0$ of 3.2 nm. The extracted values are also consistent with previous studies on Nb\textsubscript{3}Al films.\cite{kwo_superconducting_1981,tanabe_nb3aloxidepb_1983,dochev_nb3al_2008}

In addition to determining \textit{B\textsubscript{c}}, the microwire geometry enables measurement of \textit{J\textsubscript{c}}, which cannot be reliably obtained from blanket films. As shown in Supporting Information \textbf{Figure S4}, two distinct temperature-dependent behaviors were observed: one group of microwires follows a depairing-limited trend, while the second exhibits a reduced slope consistent with vortex-mediated dissipation. For wires in the depairing regime, the extracted \textit{J\textsubscript{c}(0 K)} reaches $3.32 \times 10^6$ A cm\textsuperscript{-2}. Notably, the temperature dependence of \textit{B\textsubscript{c}} remains unchanged after microwire fabrication, indicating that the microfabrication process does not introduce localized damage and that the measured depairing current reflects the superconducting properties of the entire film rather than a confined region.

\begin{table*}
    \centering
    \caption{\label{tab:table3} Superconducting characteristics of H2-AR microwires with varying lengths and widths including critical temperature, residual resistance ratio, normal resistance, sheet kinetic inductance, superfluid density, and London penetration depth. * indicates that the value was not measured.}
    \begin{tabular*}{\textwidth}{@{\extracolsep{\fill}} c c c c c c c c}
    \hline
    Wire Width ($\mu$m) & 
    Wire Length ($\mu$m) & 
    $T_c$ (K) & 
    RRR & 
    \makecell{$R_{norm}$\\($\Omega / \square$)} &
    \makecell{$L_k$ at 10 mK\\(pH/$\square$)} & 
    \makecell{$n_s$ at 10 mK\\($\times 10^{26}$ m$^{-3}$)} & 
    \makecell{$\lambda_{GL}$ at 0 K\\(nm)} \\ 
    \hline
        1 & 20 & 16.3 & 1.43 & 6.18 & 0.524 & 1.61 & 182 \\

        1 & 50 & 16.4 & 1.41 & 6.94 & 0.588 & 1.44 & 192 \\

        1 & 100 & 16.3 & * & 8.26 & 0.700 & 1.21 & 210 \\

        2 & 50 & 16.4 & * & 4.60 & 0.387 & 2.18 & 156 \\

        5 & 10 & 16.6 & 1.45 & 8.98 & 0.747 & 1.13 & 217 \\

        5 & 20 & 16.6 & 1.45 & 5.91 & 0.492 & 1.72 & 176 \\ 

        5 & 50 & 16.6 & * & 4.01 & 0.334 & 25.3 & 236 \\

        5 & 100 & 16.6 & * & 3.711 & 0.309 & 2.73 & 139 \\

        10 & 50 & 16.5 & 1.42 & 4.36 & 0.365 & 2.32 & 152 \\
        \hline
    \end{tabular*}
\end{table*}

Beyond accessing zero-temperature \textit{B\textsubscript{c}} and \textit{J\textsubscript{c}}, the transport measurements on the microwires can provide valuable information on the sheet \textit{L\textsubscript{k}} of the devices. The \textit{L\textsubscript{k}} for Nb\textsubscript{3}Al microwires can be estimated from the resistance of the normal state just above \textit{T\textsubscript{c}} following Equation \ref{eq2}:\cite{annunziata_tunable_2010}

\begin{equation}
    L_k = \frac{R_{norm}h}{2\pi^2\Delta} \frac{1}{tanh(\frac{\Delta}{2k_bT})}
    \label{eq2}
\end{equation}

\noindent where, $R_{norm}$ is the sheet resistance in the normal state, \textit{h} is the Planck constant, $\Delta$ is the BCS superconducting gap at temperature \textit{T} $<<$ \textit{T\textsubscript{c}} ($\Delta$ = 1.76 \textit{k\textsubscript{b}T\textsubscript{c}}), and \textit{k\textsubscript{b}} is the Boltzmann constant. The superfluid density (\textit{n\textsubscript{s}}) of the material can then be calculated from \textit{L\textsubscript{k}} via Equation \ref{eq3}:\cite{tinkham_michael_introduction_1996}

\begin{equation}
    n_s = \frac{m}{2e^2L_k} \frac{1}{d}
    \label{eq3}
\end{equation}

\noindent where, \textit{m} is the electron mass, \textit{e} is the fundamental charge, and \textit{d} is the film thickness.

\textbf{Table \ref{tab:table3}} summarizes the \textit{L\textsubscript{k}} and \textit{n\textsubscript{s}} values extracted for all microwire devices fabricated on the H2-AR film. The \textit{L\textsubscript{k}} values vary between 0.309 to 0.747 pH/$\square$ which are substantially lower than those typically reported for nitride superconducting thin films, where \textit{L\textsubscript{k}} can range from 30 to 200 pH/$\square$ in NbN.\cite{frasca_nbn_2023}  This difference is largely geometric: high-\textit{L\textsubscript{k}} nitride films are commonly $<10$ nm thick, whereas the Nb\textsubscript{3}Al films studied here are 210 nm thick. Because $L_k \propto 1/d$ in the thin-film limit, direct comparison of \textit{L\textsubscript{k}} across different material platforms is misleading when thickness differs by more than an order of magnitude. Assuming $L_k \propto 1/d$ continues to hold, the performance gap between Nb\textsubscript{3}Al and nitride superconductors could be narrowed.\cite{yoshida_modeling_1992} For instance, if Nb\textsubscript{3}Al can be produced with a film thickness of 10 nm, \textit{L\textsubscript{k}} could reasonably be expected to increase up to 15.7 pH/$\square$.

A more appropriate metric for cross-material comparison is the superconducting carrier density \textit{n\textsubscript{s}}, which is a bulk quantity independent of film thickness. The estimated \textit{n\textsubscript{s}} for the H2-AR microwires range from $1.13 \times 10^{26}$ to $2.7 \times 10^{26}$  m\textsuperscript{-3}  (see \textbf{Table \ref{tab:table3}}). These values were calculated using Equation 3 under the assumption that the effective mass of the charge carriers equals the free-electron mass; while this approximation is imperfect for A15 compounds, it introduces no more than $\sim 10$\% uncertainty.  For comparison, high-\textit{L\textsubscript{k}} NbN has been found to have \textit{n\textsubscript{s}} $\approx 2.194 \times 10^{26}$ m\textsuperscript{-3} , whereas elemental Nb has $n_s \approx 2.80 \times 10^{27}$ m\textsuperscript{-3} and correspondingly low \textit{L\textsubscript{k}}.\cite{dutta_superfluid_2022} The comparable \textit{n\textsubscript{s}} values of Nb\textsubscript{3}Al relative to nitrides suggest that further reductions in film thickness, while maintaining similar \textit{n\textsubscript{s}}, would substantially increase \textit{L\textsubscript{k}} of Nb\textsubscript{3}Al, potentially approaching that of nitride superconductors.

The London penetration depth can also be estimated from resistivity according to the GL theory for a dirty BCS superconductor.\cite{orlando_critical_1979,klein_calculations_1980} The assumption that Nb\textsubscript{3}Al is a dirty superconductor is made based on an estimated mean free path of 0.5 nm and coherence length similar to that found in this and other works.\cite{klein_calculations_1980} The penetration depth is given by Equation \ref{eq4}:

\begin{equation}
    \lambda_{GL}(0\; K) = 6.42 \times 10^4 \sqrt{\frac{\rho}{T_c}}
    \label{eq4}
\end{equation}

\noindent where $\rho$ is the resistivity of the normal state near \textit{T\textsubscript{c}} in units of $\Omega$ cm and $\lambda_{GL}$ is the penetration depth in nm.\cite{orlando_critical_1979} As shown in \textbf{Table \ref{tab:table3}}, extracted $\lambda_{GL}$ values for the microwires ranged from 139 nm to 217 nm. The variations result from differences in \textit{R\textsubscript{norm}}, but the values are generally comparable to the previously reported value of 210 nm.\cite{dochev_thin_2017} Since these $\lambda_{GL}$ values are comparable to the film thickness ($\sim 210$ nm), the devices operate in a regime where magnetic fields penetrate the entire film rather than being confined to the surface, leading to quasi-2D electrodynamics.\cite{tinkham_michael_introduction_1996} This is consistent with the linear \textit{B\textsubscript{c}(T)} behavior observed in \textbf{Figure \ref{fig:fig3}(b)} and supports the presence of reduced dimensionality in the superconducting state despite the nominally three-dimensional geometry.

\begin{figure}[H]
  \centering
  \includegraphics[scale=0.75]{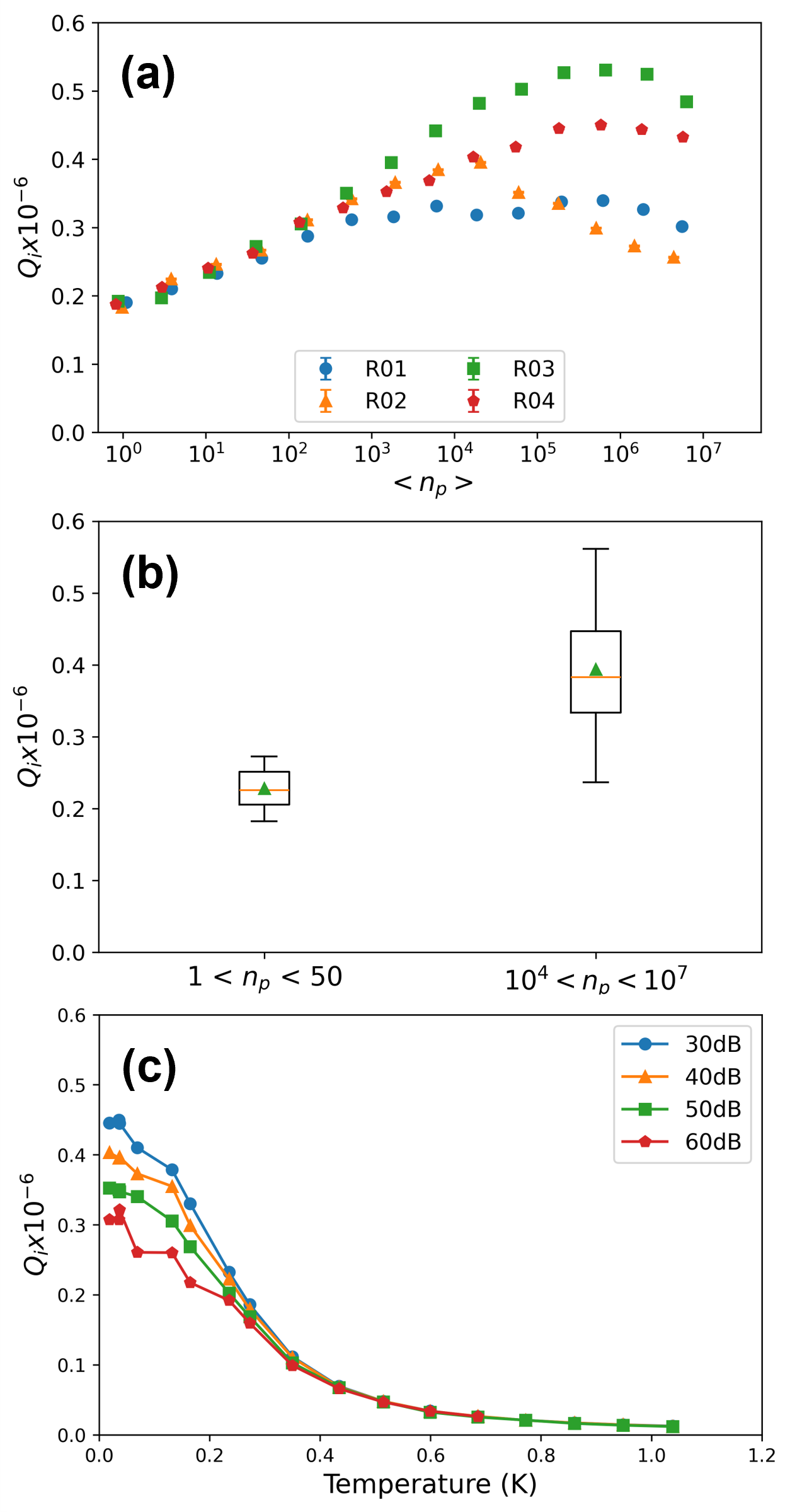}
  \caption{\textbf{Measuring microwave loss in H2-AR superconducting resonators:} (a) $Q_i$ vs microwave power expressed as the average photon number for 4 hanging CPW resonators. (b) Boxplots of $Q_i$ vs average photon number for all resonators in a set across two measurements (total of 8 measurements). The green arrow denotes the mean $Q_i$ within the set. (c) $Q_i$ vs temperature for multiple applied microwave powers.}
  \label{fig:fig4}
\end{figure}

For quantifying the microwave properties of the A15 thin films, signals were transmitted through a feedline capacitively coupled to several hanging resonators fabricated from the sample H2-AR. The CPW resonators employed in this work had 6 \textmu m wide center conductors with a 3 \textmu m gap to ground, making them sensitive to interfaces near the surface and interface losses while maintaining a characteristic impedance of 50 $\Omega$ on sapphire substrates. The transmitted frequency responses were then fit to Equation \ref{eq5} to extract the internal quality factor:\cite{megrant_planar_2012}

\begin{equation}
    S^{-1}_{21}(f) = 1 + \frac{Q_i}{|Q_c cos\phi|} \frac{e^{i\phi} }{1+2iQ_i(\frac{f-f_0}{f_0})}
    \label{eq5}
\end{equation}

\noindent where $S^{-1}_{21}$ is the inverse of the complex-valued transmission through the feedline, \textit{Q\textsubscript{i}} is the internal quality factor of the resonator, \textit{Q\textsubscript{c}} is the resonator-feedline coupling quality factor, $\phi$ is the relative phase of the transmitted signal, and \textit{f\textsubscript{0}} is the resonance frequency. Additionally, the applied power was converted to the expected number of microwave photons, \textit{n\textsubscript{p}}, in the resonator using Equation \ref{eq6}, where \textit{P\textsubscript{in}} is the input power, \textit{Q\textsubscript{t}} is the total or loaded quality factor of the resonator, and $\omega_r$ is the resonant angular frequency $(2 \pi f_r)$.\cite{bruno_reducing_2015}

\begin{equation}
    \langle n_p \rangle = \frac{2}{\hbar \omega_r^2} \frac{Q_t^2}{Q_c} P_{in}
    \label{eq6}
\end{equation}

\textbf{Figure \ref{fig:fig4}(a)} shows the \textit{Q\textsubscript{i}} for four resonators with frequencies between 4.5 and 5.25 GHz as a function of $\langle n_p \rangle$. As $\langle n_p \rangle$ increases, \textit{Q\textsubscript{i}} slowly increases up to a maximum value then begins to decrease again. The increase of \textit{Q\textsubscript{i}} at  higher values of $\langle n_p \rangle$ is consistent with the saturation of two level systems (TLS) with higher incident power. The median \textit{Q\textsubscript{i}} at the single photon limit for these resonators was found to converge at $2.26 \times 10^5$, suggesting a moderate density of TLS defects in the material. For high-quality Nb grown on Si substrates, single photon \textit{Q\textsubscript{i}} greater than $1\times 10^6$ are standard.\cite{wang_impact_2024} The observed decrease in \textit{Q\textsubscript{i}} at large $\langle n_p \rangle$ values cannot be explained by TLS’s, but rather by alternative loss channels such as quasiparticles induced by the effective heating of the resonators at relatively high applied microwave powers. This behavior is not expected for high \textit{T\textsubscript{c}} materials like Nb\textsubscript{3}Al, but can occur in superconductors such as Al which have a much smaller superconducting gap.\cite{de_visser_evidence_2014} Based on the observed behavior of these resonators, TLS losses and quasiparticle losses are in strong competition, limiting the maximum \textit{Q\textsubscript{i}} to less than $6\times 10^5$.

\textbf{Figure \ref{fig:fig4}(b)} presents the spread of \textit{Q\textsubscript{i}} from 2 separate measurements for two ranges of \textit{n\textsubscript{p}}: i) the low-power limit with \textit{n\textsubscript{p}}$<50$, approximating the single photon limit; ii) the high power limit where $10^4<n_p<10^7$, showing behavior near the maximum in \textit{Q\textsubscript{i}(n\textsubscript{p})}. Measurements were taken several hours apart and thus include effects due to time-variation of \textit{Q\textsubscript{i}}. The low power results qualitatively confirm that the low-\textit{n\textsubscript{p}} losses across the resonators are similar based on the narrow distribution of \textit{Q\textsubscript{i}} across all measurements and resonators. At higher values of \textit{n\textsubscript{p}} however, each resonator shows a different level of quasiparticle-induced loss, causing a larger spread in \textit{Q\textsubscript{i}} as the crossover between TLS saturation and quasiparticle generation occurs at a different $\langle n_p \rangle$ values, seen as changes in $n_p(Q_{i, max})$ for the various resonators shown in \textbf{Figure \ref{fig:fig4}(a)}. The high $Q_i(n_p=1)$ of Nb\textsubscript{3}Al resonators show that the material is reasonable for use in SC circuits with little change from the procedures used here. Removal of Al from the surface of films would improve the high $\langle n_p \rangle$ \textit{Q\textsubscript{i}} by reducing the nonequilibrium quasiparticle density through increased band gap.

\textbf{Figure \ref{fig:fig4}(c)} shows the temperature dependence of high-$n_p$ $Q_i$ for R04 ($f_0 = 5.23$ GHz) measured at multiple applied microwave powers (proxy variable for $n_p$), from a base temperature of $\sim 15$ mK up to $\sim 1$ K. Temperature was varied manually through this range and the resonator was allowed to thermalize at each temperature step. $Q_i$ for each input power are notably different at low temperatures, related to differences in TLS and microwave-induced quasiparticles, but decrease quickly as temperature increases. At  about 300 mK, all powers result in nearly identical $Q_i$ and further degrade at the same rate as temperature increases further. The degradation in microwave performance  is quite unexpected for a material with a $T_c>15$ K, considering that the temperature range for this measurement was below 10\% of $T_c$. On the other hand, the observed temperature dependence strongly implies that $Q_i(T)$ is dominated by thermal quasiparticle losses associated with a  low-$T_c$ phase such as Al.\cite{alexander_power_2025} This observation points to the presence of Al on the top surface of the films after the RTP step.

\begin{figure*}
  \centering
  \includegraphics[scale=0.5]{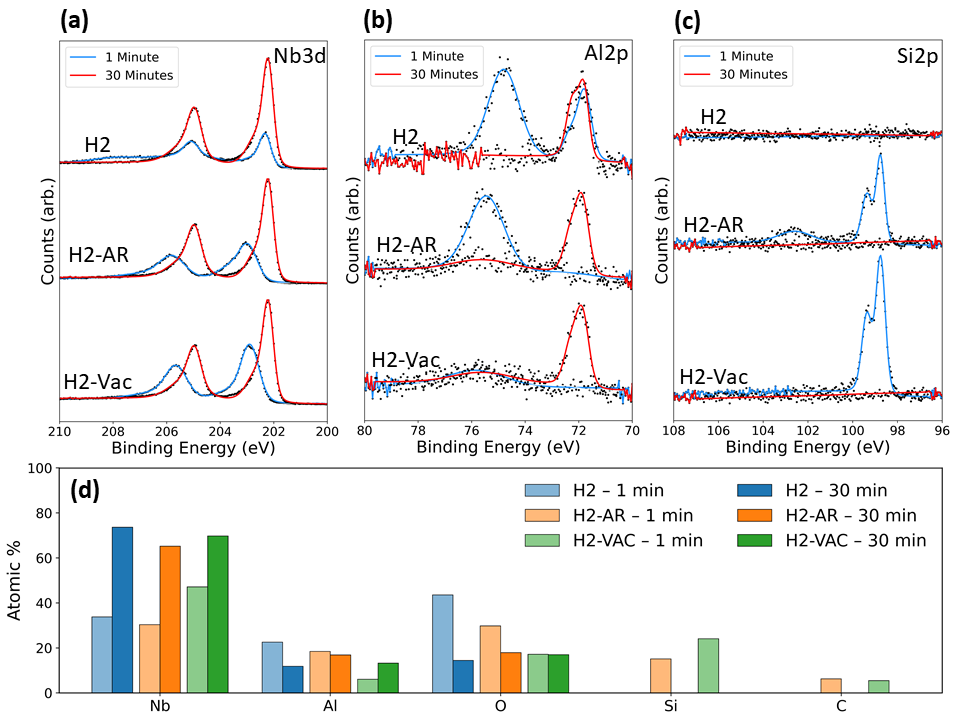}
  \caption{\textbf{XPS composition analysis for samples H2, H2-AR, and H2-VAC surface and bulk after 1 min and 30 min of Ar etching, respectively:} (a) Nb3d spectra showing Nb oxide (Nb3d\textsubscript{5/2} at 203 eV) at the surface (blue lines) of the film with metallic Nb (Nb3d\textsubscript{5/2} at 202.3 eV) dominating in the bulk (red lines). (b) Al2p spectra showing high oxidation at the surface (Al2p at 75.5 eV) of  H2 and H2-AR giving way to metallic Al in the bulk of all films (Al2p at 72 eV). (c) Si2p spectra indicating SiOx (Si2p at 103 eV) and elemental Si (Si2p doublet at 99 eV) at the surface of films only after RTP. (d) Elemental composition of the three films after 1 minute (light shade) and 30 minutes (dark shade) of Ar sputtering using peak areas normalized by RSF.}
  \label{fig:fig5}
\end{figure*}

Motivated by the indications of surface-related losses observed in the superconducting resonator measurements, we performed XPS analysis on H2, H2-VAC, and H2-AR films. \textbf{Figure \ref{fig:fig5}} shows the XPS spectra and elemental analysis results collected after 1 minute and 30 minutes of in-situ Ar ion etching. The 1-min Ar etch removes adventitious carbon allowing analysis of the near-surface region ($\sim 2$ nm depth), whereas 30-min etch  probes the bulk of the films. Initial survey scans of samples following RTP revealed Nb, Al, C, and O, as expected, along with moderate amounts of Si. High-resolution scans of the Nb 3d, Al 2p, Si 2p, O1, and C1s were then acquired, and peak areas were then normalized using Scofield relative sensitivity factors (RSFs), normalized to C1s, to determine the elemental composition.\cite{scofield_theoretical_1972} 

\textbf{Figure \ref{fig:fig5}(a)} shows the Nb 3d spectra for H2, H2-AR, and H2-VAC after 1 minute and 30 minutes of Ar ion etch. The Nb3d transition consists of a spin-orbit doublet with a spacing of approximately 2.7 eV, such that each oxidation state contributes two well-resolved peaks to the spectrum. For simplicity, oxidation states will be distinguished by their 3d\textsubscript{5/2} binding energy (BE). In the as-deposited H2 sample, there is a small amount of Nb\textsubscript{2}O\textsubscript{5} (BE of 207 eV) present on largely metallic Nb (BE of 202.2 eV). This is consistent with a thin layer of Nb\textsubscript{2}O\textsubscript{5} formed on the as-deposited films as a result of extended air exposure. On the other hand, H2-AR and H2-VAC films that underwent the RTP process, Nb3d peaks shift to larger  BEs, which is consistent with formation of NbO and/or NbC on the surface which are indistinguishable as both have a Nb3d\textsubscript{5/2} peaks near 203 eV. After 30 min of etching, the spectra from all samples show metallic Nb peaks, indicating that the surface oxygen and carbon does not diffuse through the film during the RTP to form oxides/carbides within the bulk. Since oxides and carbides are hosts for TLS’s and are present only at the surface of the samples, etching of the surface to remove the lossy material would serve to increase $Q_i$ in the single photon limit.

\textbf{Figure \ref{fig:fig5}(b)} shows the Al 2p spectra for the same samples shown in \textbf{Figure \ref{fig:fig5}(a)}. In the as-deposited H2 film, signals from both metallic and oxidized Al (AlOx with BE $\approx 74.5$ eV and Al\textsubscript{2}O\textsubscript{3} with BE $\approx 75.5$ eV) are visible with the oxide peak being dominant. This is consistent with a thin oxide layer over the as-desposited Nb\textsubscript{3}Al film. Similar to the Nb3d spectra, however, the surface of H2-AR sample is almost wholly covered by Al\textsubscript{2}O\textsubscript{3} with no discernable metallic peak. This can be ascribed to the oxidation during the Ar annealing process by the background O present within the RTP chamber. While significantly weaker in signal intensity, H2-VAC, also shows an  Al 2p peak with BE$=75.5$ eV corresponding to Al\textsubscript{2}O\textsubscript{3} presence on its surface. However, comparing the signal intensities between H2, H2-AR, H2-VAC confirms that vacuum annealing depletes the Al content near the top surface. Moreover, the Al2p spectra for all samples after 30 min of Ar etching are dominated by metallic Al, though a very slight oxide peak is visible in the RTP-annealed H2-AR and H2-VAC samples. Therefore, while not obvious from the Nb3d spectra, the Al2p results point to small amounts of O diffusing through the Nb\textsubscript{3}Al films forming Al\textsubscript{2}O\textsubscript{3}-containing regions within the bulk. Partial oxidation of the Al suggests a high TLS density throughout the film which limits the $Q_i$ observed for the resonators at low $\langle n_p \rangle$.

Alongside  Nb and Al, and the expected surface contaminants O and C, our XPS measurements revealed moderate amounts of Si on the Nb-Al films surfaces following RTP. The Si 2p spectra for the two RTP-annealed samples, H2-AR, and H2-VAC, are shown in \textbf{Figure \ref{fig:fig5}(c)} and compared to that of the as-deposited H2 sample. Only the as-deposited sample shows no detectable Si on its surface, indicating that Si incorporation occurs during the RTP. In the Ar-annealed sample (H2-AR), most of the Si appears in the metallic state, though a small SiO2 component is present. RTP under high vacuum reduces the SiO\textsubscript{2} contribution without substantially affecting the metallic peak intensity. The two peaks observed in the metallic range arise from the Si 2p spin-orbit doublet resolved under low pass energy, and should not be interpreted as distinct oxidation states. After 30 min of in-situ Ar etching, all Si-related signals disappear, indicating that Si contamination is limited to the top surface of the films. Because Si incorporation is evident only after RTP, it is assumed that the presence of Si is due to contamination of the RTP chamber by other processes. XRD patterns show Si forms compounds with Nb, contributing to disorder, but microwave losses associated with Si-containing phases have not been isolated from other loss channels. Si incorporation can be reduced by heat treatment in a cleaner (ultra) high vacuum environment.

After examination of the Nb 3d, Al 2p, and Si 2p spectra individually, we consider the full elemental composition obtained from the Nb, Al, O, C, and Si peak areas corrected by the RSF (\textbf{Figure \ref{fig:fig5}(d)}). As seen in \textbf{Figure \ref{fig:fig5}(d)}, the Nb:Al ratio of the 1-min etched surface region changes markedly across the three films, moderately increasing from 1.48:1 in the as-deposited H2 sample to 1.67:1 in H2-AR and increasing drastically to 7.83:1 in H2-VAC. In contrast, the Nb:Al ratio in the bulk varied from 6.17:1 in as-deposited H2 to 3.82:1 and 5.38:1 in H2-AR and H2-VAC, respectively. This trend indicates that RTP removes Al from the surface of the films, but that the effect is much more drastic when RTP occurs in a high vacuum environment. The excess Al present at the top surface of H2-AR, seen as nanopillars in AFM topographs (see Supporting Information, \textbf{Figure S3}), would be present on the resonator devices, contributing to quasiparticle losses at high applied microwave powers and elevated temperatures.

Regarding contaminants, the Si detected after RTP is confined to the surface; as seen both in \textbf{Figure \ref{fig:fig5}(c) and \ref{fig:fig5}(d)} all Si signals disappear after 30-min in-situ Ar etching, confirming that it does not diffuse into the bulk. Oxygen concentrations are similar in the bulk of all samples and are interpreted as physisorbed species introduced during ion etch or analysis steps rather than incorporated oxides, consistent with minimal Nb or Al oxidation features in \textbf{Figure \ref{fig:fig5}(a) and \ref{fig:fig5}(b)}. Carbon is present only on the annealed films and persists after the 1-min etch, despite removal of adventitious carbon; this suggests carbide formation during RTP, potentially arising from the same contamination pathway as the observed Si. These observations suggest that the Si- and C-related surface contamination originates from the RTP environment rather than the sputtering process, and could be mitigated by the use of alternative susceptor materials or cleaner post-deposition annealing processes.

\section{Conclusion}
In this work, we demonstrated the growth of A15-phase Nb\textsubscript{3}Al thin films by magnetron co-sputtering and showed that ex-situ rapid thermal processing at 1000 $^\circ$C increases the superconducting transition temperature to as high as 16.4 K. Microwire devices patterned from these films revealed superconducting parameters consistent with high-performance A15 materials, including a zero-temperature coherence length of 3.1 nm, $B_c(0) = 33.2$ T, $J_c = 3.2 \times 10^6$ A cm\textsuperscript{-2}, superfluid densities on the order of 1026 m\textsuperscript{-3}, and London penetration depths between $135- 220$ nm. Coplanar waveguide resonators fabricated on Nb\textsubscript{3}Al exhibited microwave frequencies between 4.5–5.25 GHz and single-photon $Q_i$ values of $2.26 \times 10^5$, demonstrating the compatibility of this material with superconducting circuit architectures. 

Although not yet competitive with the best-performing nitride superconductors, these results establish Nb\textsubscript{3}Al as a promising platform for superconducting electronics, including potential applications in quantum devices. Improving surface chemistry, particularly the removal of Al-rich oxides and RTP-induced contaminants, represents a clear pathway toward enhanced device performance. Future efforts will focus on refining annealing and cleaning protocols, exploring alternative annealing processes, and investigating device geometries below the London penetration depth to further leverage the intrinsic properties of A15 Nb\textsubscript{3}Al.


\begin{acknowledgments}
J.F. and K.S. acknowledge Leonard Feldman and Hussain Hijazi at Rutgers University for conducting the Rutherford Backscattering measurements. J.F. and K.S. also acknowledge Tom Hazard of MIT Lincoln Laboratory and Alan Kramer of Laboratory for Physical Sciences for valuable discussions.
\end{acknowledgments}

\bibliographystyle{apsrev4-1}
\bibliography{Nb3Al_bib}

\clearpage
\onecolumngrid       
\appendix  

\section*{Supporting Information}

\setcounter{figure}{0}
\renewcommand{\thefigure}{S\arabic{figure}}

\input{Nb3Al_SI}

\end{document}

%% file: Nb3Al_SI.tex


\makeatletter
\makeatother

\DeclareLanguageAlias{en}{english}

\addto\captionsenglish{\renewcommand{\figurename}{\textbf{Figure}}}
\renewcommand{\thefigure}{\textbf{S\arabic{figure}}}

\draft 



\title{Supporting Information\\An evaluation of A15 Nb\textsubscript{3}Al superconducting thin films for application in quantum circuits}

\author{Joseph Falvo}
\affiliation{Department of Materials Science and Engineering, University of Maryland, College Park, Maryland 20740, USA}
\affiliation{Laboratory for Physical Sciences, University of Maryland, College Park, Maryland 20740 USA}

\author{Brooke Henry}
\affiliation{Department of Physics, University of Colorado, Boulder, Colorado 80309, USA}

\author{Bernardo Langa Jr.}
\affiliation{Laboratory for Physical Sciences, University of Maryland, College Park, Maryland 20740 USA}
\affiliation{Department of Physics, University of Maryland, College Park, Maryland 20740, USA}

\author{Rohit Pant}
\affiliation{Laboratory for Physical Sciences, University of Maryland, College Park, Maryland 20740 USA}
\affiliation{Department of Physics, University of Maryland, College Park, Maryland 20740, USA}

\author{Ashish Alexander}
\affiliation{Laboratory for Physical Sciences, University of Maryland, College Park, Maryland 20740 USA}
\affiliation{Department of Physics, University of Maryland, College Park, Maryland 20740, USA}

\author{Jason Dong}
\affiliation{Laboratory for Physical Sciences, University of Maryland, College Park, Maryland 20740 USA}
\affiliation{Department of Physics, University of Maryland, College Park, Maryland 20740, USA}

\author{Kasra Sardashti}
\altaffiliation{Corresponding author: ksardash@umd.edu}
\affiliation{Laboratory for Physical Sciences, University of Maryland, College Park, Maryland 20740 USA}
\affiliation{Department of Physics, University of Maryland, College Park, Maryland 20740, USA}

\date{\today}

\maketitle 

\begin{figure} [H]
\centering
\includegraphics[scale=0.75]{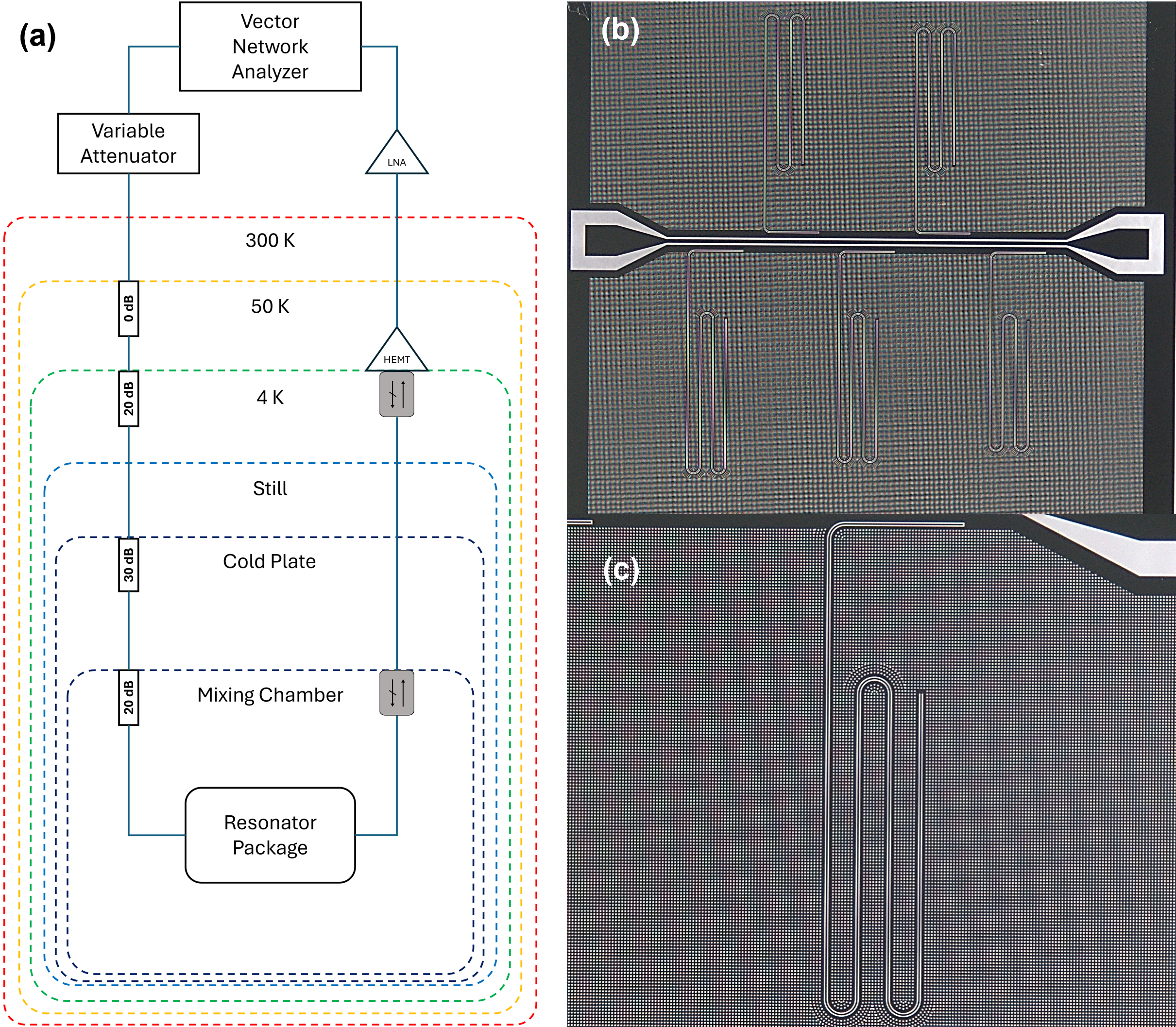}
\caption{\label{fig:FIGS1}\textbf{Resonator measurement setup:} (a) Dilution refrigerator wiring diagram for resonator measurements showing attenuators, circulators, and HEMT amplifiers, as well as VNA and variable attenuator used to minimize noise induced by changing output power of VNA. (b) Optical microscope image of a CPW resonator chip consisting of 5 resonators capacitively coupled to a common microwave feedline. (c) Optical microscope image of single hanger resonator.}
\end{figure}

\newpage

\begin{figure}[H]
\centering
\includegraphics[scale=0.3]{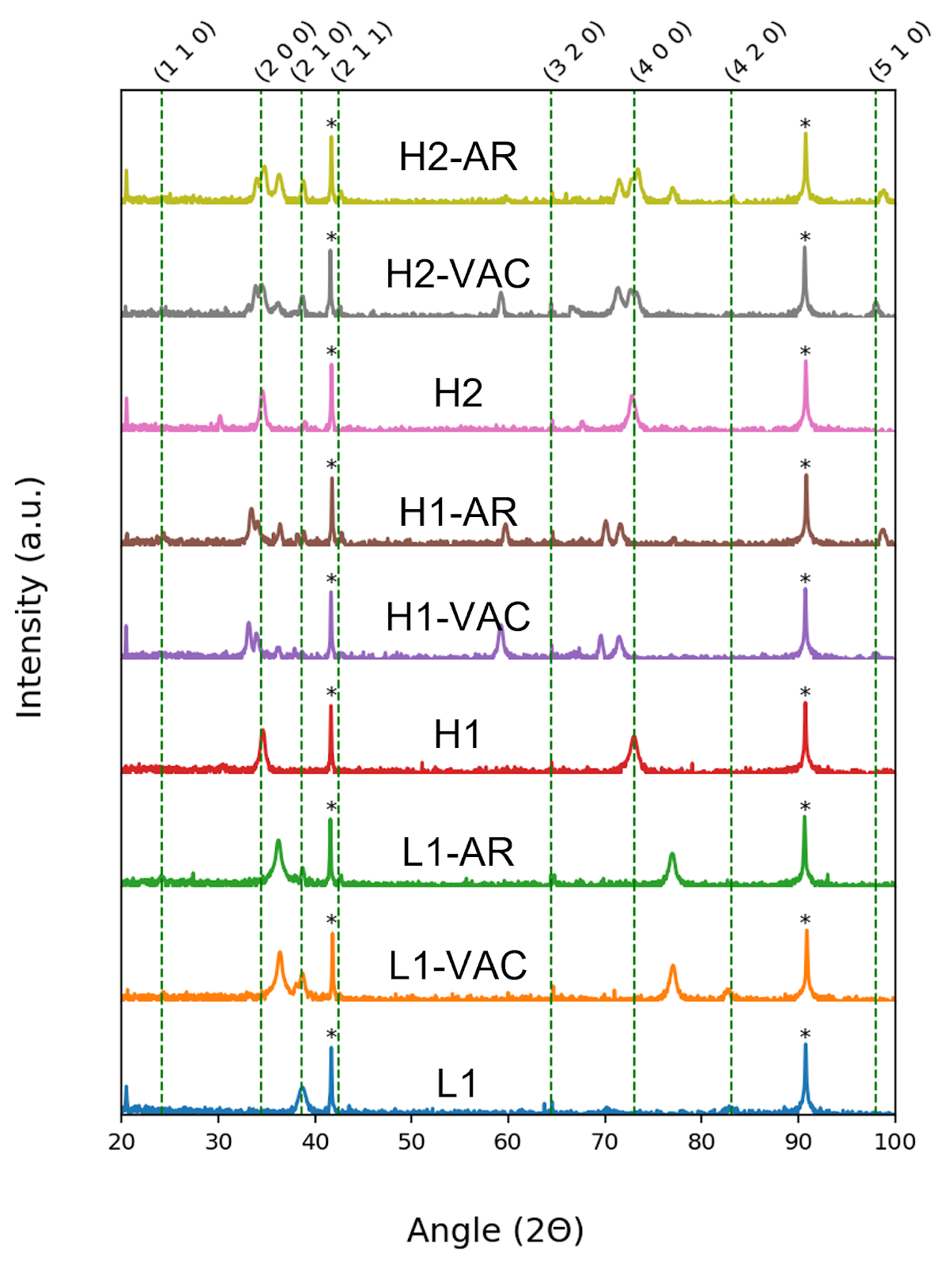}
\caption{\label{fig:FIGS2} \textbf{XRD patterns for all samples presented in this work:} * denotes peaks related to the sapphire substrate while green dashed lines indicate peak positions from A15 Nb3Al. Unlabeled peaks are attributed to carbide phases and Nb-Al-Si ternaries, though exact phases are undetermined.}
\end{figure}

\newpage

\begin{figure} [H]
\centering
\includegraphics[scale=0.3]{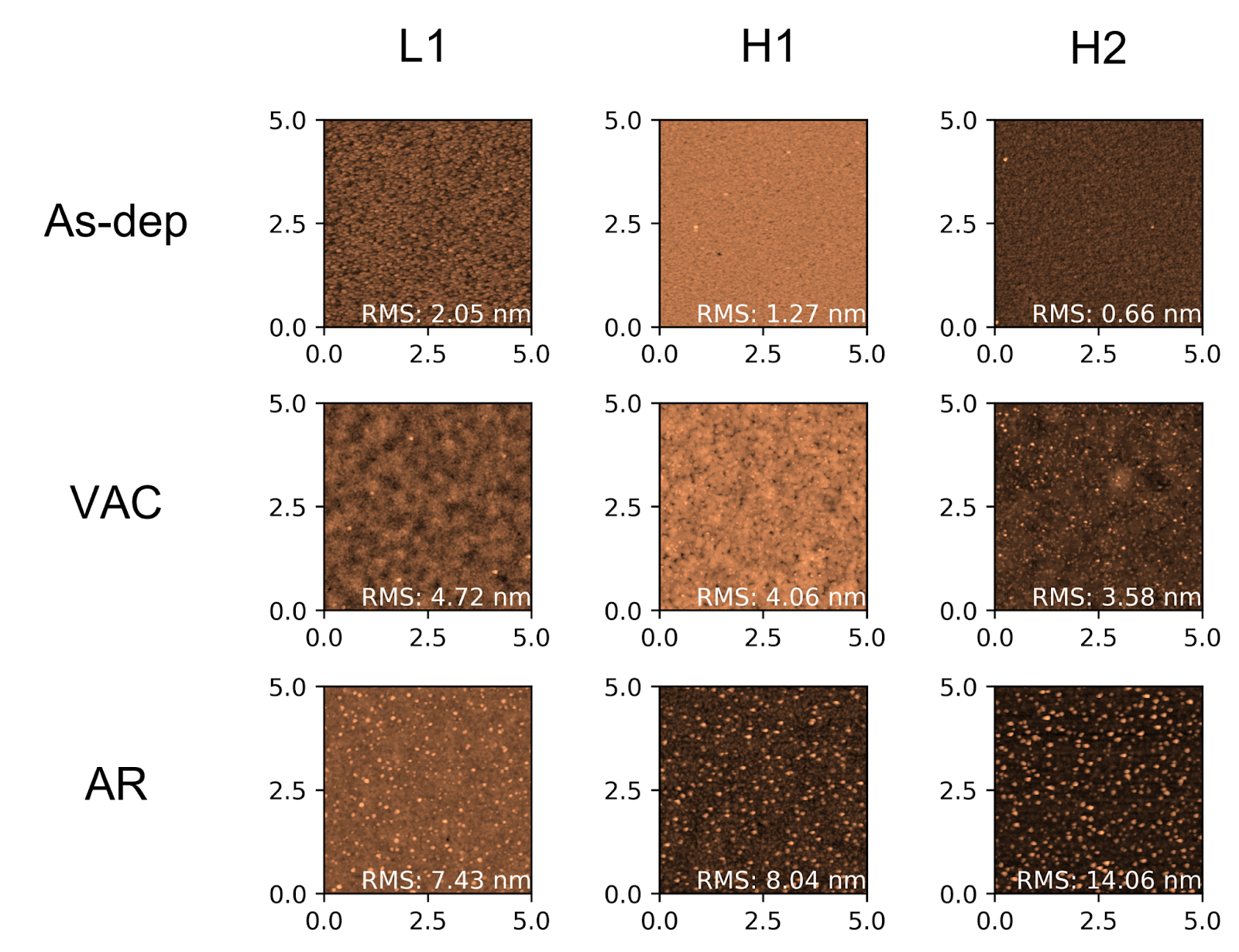}
\caption{\label{fig:FIGS3} \textbf{AFM surface topography maps  for all samples:} Roughness from vacuum RTP originates from “pits” formed by Al leaching. Roughness in AR samples occurs as pillars on the surface up to 80 nm tall in H2-AR. Both processes require diffusion of Al towards the surface. RMS roughnesses are shown in the lower right corner of each map.}
\end{figure}

\newpage

\begin{figure} [H]
\centering
\includegraphics[scale=0.6]{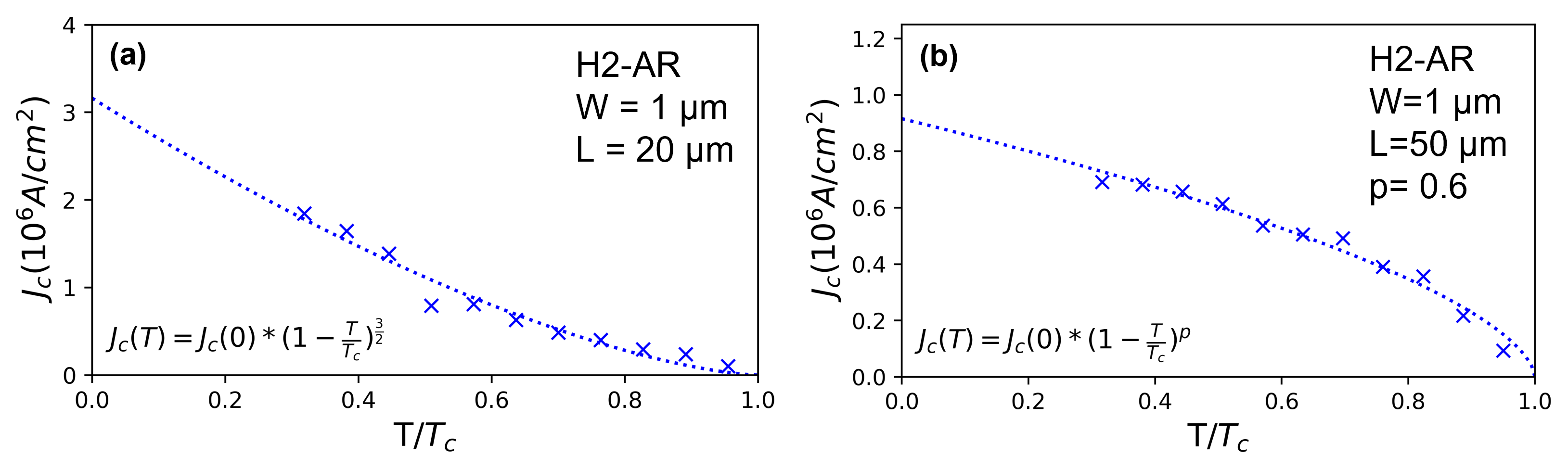}
\caption{\label{fig:FIGS4}$J_c(T)$ for (a) 1 \textmu m $\times$ 20 \textmu m wire fit to GL depairing behavior. (b) 1 \textmu m $\times$ 50 \textmu m wire fit to a power law with exponent 0.6. In case (a), current is limited by depairing events, likely at rough edges or thin portions of the wire. Case (b) has $J_c$ limited by weak links or Josephson-like behavior at some point along the length. In neither case did the measured $J_c$ approach the maximum limit from the penetration depth and critical field.}
\end{figure}

\newpage


%
%

%


%